\documentclass[paper,12pt]{article}

\newcommand{\sect}[1]{\setcounter{equation}{0}\section{#1}}
\begin{document}
\topmargin 0pt \oddsidemargin 0mm

\renewcommand{\thefootnote}{\fnsymbol{footnote}}
\begin{titlepage}

\vspace{2mm}
\begin{center}
{\Large \bf Vacuum Structure of de Sitter Space}
 \vspace{12mm}

{\large Sang Pyo Kim\footnote{e-mail address: sangkim@kunsan.ac.kr}}\\

\vspace{5mm}
 {\em Department of Physics, Kunsan National University, Kunsan 573-701,
 Korea}
\date{\today}
\vspace{5mm}\\
{(Dated: August 3, 2010)}

\end{center}

\vspace{10mm} \centerline{{\bf{Abstract}}} \vspace{5mm}
In the in-/out-state formalism we find the exact one-loop effective action of a massive scalar
field in the global coordinates of de Sitter spaces, which is a gravity analog of the Heisenberg-Euler
action in QED. The nonperturbative effective action, modulo the angular momentum sum, has an imaginary part in all even dimensions, but the imaginary part of the effective action is zero in all odd dimensions. However, 
in the zeta-function regularization for angular momenta, the weak-curvature expansion of the renormalized effective action vanishes in any even dimension, while the real part is finite in any odd dimension. This implies that de Sitter spaces may be stable against particle production at one-loop.
\end{titlepage}

\newpage
\renewcommand{\thefootnote}{\arabic{footnote}}
\setcounter{footnote}{0} \setcounter{page}{2}

\sect{Introduction}

Recently Polyakov has argued that any even dimensional de Sitter space (dS) cannot be an ``external
manifold'' due to the instability from Schwinger mechanism for cosmic particle production, the so-called
``cosmic laser'' \cite{polyakov}. That the dS space emits a thermal spectrum just as black holes do the Hawking radiation was first shown by Gibbons and Hawking \cite{gibbons-hawking}.
The dS radiation is the gravity analog of Schwinger pair production by a strong electric field \cite{schwinger,heisenberg-euler} and has the Unruh temperature, the Hubble constant modulo $2 \pi$.
This cosmic laser may raise a serious challenge to cosmologists since the present accelerating universe is driven either by dark energy or by a cosmological constant. The Einstein gravity with the positive cosmological constant is the dS space.

The dS space has played a special role for several decades because it has the dS group, the maximal spacetime symmetry for a given dimension. The dS group makes free fields separable and solvable, and thus constructs quantum fields in dS space (for review, see \cite{birrel-davies}). A different coordinate system for dS space leads to a different set of field modes, which in turn defines a vacuum. Hence, which vacuum is more physical than others and whether vacua are stable or not have been an issue of continual debates. The dS space has one-parameter family of vacua invariant under the dS group \cite{chernikov-tagirov,mottola,allen}, and one vacuum is a squeezed state of the other \cite{BMS}.
The selection of the vacuum requires some guiding principles, one of which is the composition
principle by Polyakov for causality of the propagator. It is not the in-/in-state (Schwinger-Keldysh) formalism  based on the Bunch-Davies vacuum but the in-/out-state formalism based on the vacua without any particle at the past and future infinities that complies with the composition principle in dS space \cite{polyakov}.

The purpose of this paper is first to find the effective action of a massive scalar field in the global coordinates of dS spaces in the in-/out-state formalism and then to discuss the physical implications in cosmology. The in-vacuum and the out-vacuum do not contain any particle incoming from the past and future infinities, respectively. In Refs. \cite{mottola,BMS,JMP} field modes for a massive scalar field that define the in- and the out-vacua are found in the global coordinates of dS spaces. In the in-/out-state formalism,
the effective action is the $S$-matrix between the out-vacuum and the in-vacuum \cite{schwinger51b} (for review, see \cite{birrel-davies})
\begin{eqnarray}
e^{i W_{\rm eff}} = \langle {\rm out} \vert {\rm in} \rangle.
\end{eqnarray}
As the in-/out-vacua are related through the Bogoliubov transformation, the effective action can be expressed
in terms of the Bogoliubov coefficients \cite{dewitt,nikishov}. In fact, the effective action is the sum of
the logarithmic function of the Bogoliubov coefficient for each mode. Using the gamma function technique \cite{KLY08,KLY10}, we are able to find the exact one-loop effective action of a massive scalar field in dS space of any dimension.

The effective action is the gravity analog of Heisenberg-Euler effective action in a constant electric field
\cite{heisenberg-euler}. These effective actions are nonperturbative in that the imaginary part describing the decay rate cannot be obtained by summing a finite number of Feynman diagrams. Instead, these effective actions sum all one-loop Feynman diagrams with even number of external legs for photons in quantum electrodynamics (QED) and gravitons in dS space. The effective action in the in-/out-state formalism should be distinguished from early works in the in-/in-state formalism. For instance, the effective actions in dS space in Refs. \cite{candelas-raine,dowker-critchley}, which are obtained using the Feynman propagator, are the result in the in-/in-state formalism. In the in-/in-state formalism the effective action, being real, does not have an imaginary part and thus cannot explain the particle production. An interesting observation by Das and Dunne is that in strong contrast with the electromagnetic duality in QED, the duality does not hold between the effective action of a dS space and that of an anti-de Sitter (AdS) space since the Feynman propagator for AdS space cannot continue analytically to that for dS space due to different boundary conditions \cite{das-dunne}. There is the gauge-gravity relation between the scalar QED effective action in a maximally symmetric $4n$-dimensional electromagnetic field and the spinor effective action in a $2n$-dimensional AdS space \cite{basar-dunne}.

The organization of this paper is as follows. In Sec. 2, we briefly review the effective action in
the in-/out-state formalism. In Sec. 3, we obtain the effective action of a massive scalar field
in dS spaces and compare it the Heisenberg-Euler effective action in scalar QED.
In Sec. 4, we propose a regularization scheme based on the zeta function and obtain
the renormalized effective action. Finally we discuss the physical implications to cosmology in conclusion.

\section{The In-/Out-State Formalism Revisited}

We revisit the effective action in the in-/out-state formalism and
then focus on dS spaces. In the in-/out-state formalism, one assumes the vacua at
the past and future infinities, which are not necessarily identical to each other.
The question of how to select the in-/out-vacua will be addressed in detail
for the dS space in Sec. 3. The vacua are assumed to exist in two regions.
The in-vacuum does not contain any particle and/or antiparticle coming from the past infinity:
\begin{eqnarray}
a_{{\rm in}, L} \vert {\rm in} \rangle = 0, \quad b_{{\rm in}, L} \vert {\rm in} \rangle = 0,
\end{eqnarray}
where $L = (l, \cdots)$ collectively denotes all quantum numbers associated with
a quantum field. Likewise, the out-vacuum does not have any particle
and/or antiparticle coming from the future infinity:
\begin{eqnarray}
a_{{\rm out}, L} \vert {\rm out} \rangle = 0, \quad b_{{\rm out}, L} \vert {\rm out} \rangle = 0.
\end{eqnarray}
Then the Bogoliubov transformations,
\begin{eqnarray}
a_{{\rm out}, L}= \mu_L a_{{\rm in}, L} + \nu_L^* a_{{\rm in}, L}^{\dagger}, \label{bog tran}
\end{eqnarray}
relate the out-states with the in-states, and vice versa. Here, field modes are assumed to decouple
for the sake of simplicity, though the mode-mixing case may be handled
in a similar but complicated manner \cite{dewitt}. The coefficients satisfy the Bogoliubov
relations
\begin{eqnarray}
| \mu_{L} |^2 - | \nu_{L}|^2 = 1.
\end{eqnarray}

For each quantum $L$ the Bogoliubov transformation may be written as a unitary transformation
\begin{eqnarray}
a_{{\rm out}, L} = U_{L} a_{{\rm in}, L} U^{\dagger}_{L}, \quad b_{{\rm out}, L} =
U_{L} b_{{\rm in}, L} U^{\dagger}_{L}.
\end{eqnarray}
The out-vacuum
\begin{eqnarray}
\vert {\rm out} \rangle = \prod_{L} U_{L} \vert {\rm in} \rangle
\end{eqnarray}
is the squeezed vacuum of the in-vacuum, since $U_L$ is either a two-mode squeeze operator for QED or a
one-mode squeeze operator for dS space, whose explicit form is given in Ref. \cite{KLY08}.
One may directly show \cite{KLY10}
\begin{eqnarray}
\vert {\rm out} \rangle =  \prod_{L}
\Biggl[ \frac{1}{\mu_{L}} \sum_{n_{L} = 0}^{\infty} \Bigl(- \frac{\nu^*_{L}}{\mu_{L}}
\Bigr)^{n_{L}} \vert n_{L}, \bar{n}_{L}; {\rm in}
\rangle \Biggr].
\end{eqnarray}
Here, $\vert n_{L}, \bar{n}_{L}; {\rm in} \rangle  = (a^{\dagger}_{{\rm in}, L} b^{\dagger}_{{\rm in}, L})^{n_L} \vert {\rm in} \rangle/n_L!$ is the multi-particle and anti-particle state.
In the in-/out-state formalism, the exact one-loop effective action ${\cal L}_{\rm eff}$
per unit volume and per unit time is given by the $S$-matrix as \cite{dewitt,nikishov}
\begin{eqnarray}
W_{\rm eff} = \int dt d^d x {\cal L}_{\rm eff} = - i \ln (\langle {\rm out} \vert {\rm in}
\rangle) = i (V T) \sum_L \ln (\mu_L^*), \label{eff act}
\end{eqnarray}
where $d$ denotes the space dimensions, $V$ and $T$ are the volume and the duration.
Using the gamma function regularization, the exact one-loop QED effective actions are
successfully worked out for time-dependent electric fields \cite{KLY08} and spatially localized electric fields \cite{KLY10}.
The vacuum persistence, which is the probability for the in-vacuum to remain in the out-vacuum, is
\begin{eqnarray}
|\langle {\rm out} \vert {\rm in} \rangle |^2 = e^{ -
 \int dt d^d x \sum_{L}
\ln (1 + |\nu_L|^2)}.
\end{eqnarray}
Here,
\begin{eqnarray}
\bar{\cal N}_L = |\nu_L|^2
\end{eqnarray}
is the mean number of produced particles with quantum number $L$ per unit volume and per unit time.
One can show the general relation between the imaginary part of
the effective action and the total mean number of produced particles \cite{KLY08,KLY10,GGT,hwang-kim}
\begin{eqnarray}
2 {\rm Im}({\cal L}_{\rm eff}) =  \sum_{L} \ln (1 +  {\cal N}_{L}). \label{gen-rel}
\end{eqnarray}
Note that the effective action (\ref{eff act}) and the general
relation (\ref{gen-rel}) should be regularized to the renormalized ones because the number of states allowed is infinite in general,
as will be shown in Sec. 4.

\section{Effective Action of dS Space}

In this section we study the in- and out-vacua of a massive scalar field in dS spaces
and then find the effective action according to Sec. 2.
A $(d+1)$-dimensional dS space has the global coordinates with the metric [in units of $c = \hbar = 1$]
\begin{eqnarray}
ds^2 = - dt^2 + \frac{\cosh^2 (Ht)}{H^2} d \Omega_d^2, \label{dS met}
\end{eqnarray}
which has the topology $R \times S_d$ and a constant scalar curvature $R = d(d+1) H^2$.
As the dS space is maximally symmetric, the massive scalar field is completely integrable.
Indeed, the massive scalar field $\Phi$ with mass $m$ is decomposed
by the spherical harmonics $Y_{L} (\Omega_d)$ of the Laplace operator on $S_d$,
\begin{eqnarray}
\Delta^{(d)} Y^{(d)}_{L} (\Omega_d) = - L^2 Y^{(d)}_{L} (\Omega_d),
\end{eqnarray}
with the eigenvalues and degeneracies \cite{rubin-ordonez}
\begin{eqnarray}
L^2 = l (l + d -1), \quad D_l^{(d)} = \frac{(l + d-2)!}{l!(d-1)!}
(2l + d-1), \quad (l = 0, 1, \cdots). \label{sph har}
\end{eqnarray}
Then the scalar field is quantized as
\begin{eqnarray}
\Phi (t, \Omega_d) =
\sum_{L} [ a_{L} \varphi(t, \Omega_d) + a^{\dagger}_{L} \varphi^*(t, \Omega_d)]. \label{quant field}
\end{eqnarray}
Here, the field mode
\begin{eqnarray}
\varphi(t, \Omega_d) = \frac{1}{\cosh^{\frac{d}{2}} (Ht)} \phi_{L} (t) Y_L (\Omega_d),
\end{eqnarray}
is determined by the solution to the one-dimensional scattering problem
\begin{eqnarray}
\Biggl[ \frac{d^2}{d t^2}  + H^2 \Bigl(\gamma^2 +
\frac{L^2 + \frac{d}{2}(\frac{d}{2}-1)} {\cosh^2 (Ht)} \Bigr) \Biggr]
\phi_{L} (t) = 0, \quad \gamma = \sqrt{\frac{m^2}{H^2} -
\frac{d^2}{4}}. \label{mode eq}
\end{eqnarray}
In this paper we consider only the massive scalar $(m > \frac{dH}{2})$, and will study light scalar $(m < \frac{dH}{2})$ or massless scalar and/or graviton elsewhere. Following Ref. \cite{mottola} for $d =3$ and Ref. \cite{BMS} for any $d$, we choose the solution given by the hypergeometric function
\begin{eqnarray}
\phi_{L} (t) = \frac{2^{l+\frac{d}{2}}}{\sqrt{2 H \gamma}}
\cosh^{l+\frac{d}{2}} (Ht) \exp \Bigl[(l+\frac{d}{2} - i \gamma )Ht
\Bigr] F \Bigl(l+\frac{d}{2}, l+\frac{d}{2} -i \gamma, 1 - i \gamma,
- e^{2 Ht} \Bigr). \label{in sol}
\end{eqnarray}
The solution has the asymptotically positive frequency
\begin{eqnarray}
\phi_{L} (- \infty ) = \frac{e^{-i \gamma H t}}{\sqrt{2 \gamma H}}. \label{asym sol}
\end{eqnarray}

The quantum field (\ref{quant field}) with the field mode (\ref{in sol})
prescribes the in-vacuum, which does not contain any particle
from the past infinity $(t = - \infty)$, since each mode
has the asymptotically positive frequency (\ref{asym sol}). Similarly, the out-vacuum is defined
with respect to field modes with the asymptotically positive frequency
\begin{eqnarray}
\phi_{L} (\infty ) = \frac{e^{-i \gamma H
t}}{\sqrt{2 \gamma H}}.
\end{eqnarray}
As the solution (\ref{in sol}) splits into one branch of positive frequency
and another branch of negative frequency at $t = \infty$, the particle annihilation
and creation operators for the in-vacuum and the out-vacuum are related through
the Bogoliubov transformations,
as explained in Sec. 2. The Bogoliubov coefficients are given by
\begin{eqnarray}
\mu_{L} &=& \frac{\Gamma (1 - i \gamma) \Gamma (- i \gamma)}{\Gamma
(l + \frac{d}{2} - i \gamma) \Gamma (1- l - \frac{d}{2} - i
\gamma)},
\nonumber\\
\nu_{L} &=& \frac{\Gamma (1 - i \gamma) \Gamma (i \gamma)}{\Gamma (l
+ \frac{d}{2}) \Gamma (1 - l -\frac{d}{2})}, \label{bog coef}
\end{eqnarray}
which satisfy the relations
\begin{eqnarray}
|\mu_{L}|^2 - |\nu_{L}|^2 = 1.
\end{eqnarray}
The mean number of produced scalar particles is
\begin{eqnarray}
\overline{\cal N}_{L} = \vert \nu_{L} \vert^2 = \Biggl(\frac{\sin \pi
(l + \frac{d}{2})}{\sinh \pi \gamma} \Biggr)^2. \label{mean num}
\end{eqnarray}
Note that the mean number vanishes for any odd-dimensional dS space $(d = {\rm even~integer})$ as pointed out in Ref. \cite{BMS}.
That particles are not produced in odd dimensional dS spaces is the consequence of a reflectionless soliton solution by the KdV equation (\ref{mode eq}) in odd dimensions \cite{polyakov}.
A semiclassical explanation is that the double Stokes lines for the Wentzel-Kramers-Brillouin actions for pair production interfere constructively in even dimensions but destructively in odd dimensions \cite{kim10}, implying no particle production.

Following the method in Refs. \cite{KLY08,KLY10}, we substitute the Bogoliubov coefficients (\ref{bog coef}) into Eq. (\ref{eff act}), use the gamma function \cite{gamma}, and then obtain the exact one-loop effective action
per unit volume and per unit time
\begin{eqnarray}
{\cal L}^{(d+1)}_{\rm eff} (H) = i \frac{\Gamma (\frac{d+1}{2})}{(2\pi)^{\frac{d+1}{2}}} mH^d \sum_{l =
0}^{\infty} D^{(d)}_l
 \int_0^{\infty} \frac{ds}{s} \frac{e^{- i \gamma s}}{1 - e^{-s}} \Biggl[1
+ e^{-s} - e^{-(l + \frac{d}{2}) s} - e^{(l+\frac{d}{2}-1)s} \Biggr].
\label{ds eff act}
\end{eqnarray}
Here, we have divided the action by the Hubble volume $
2\pi^{(d+1)/2}/\Gamma (\frac{d+1}{2})H^d$ and the Compton time $1/m$.
Finally, after doing the contour integral along a
quarter circle of infinite radius in the fourth quadrant, we
obtain the effective action
\begin{eqnarray}
{\cal L}^{(d+1)}_{\rm eff} (H) &=&  \frac{\Gamma (\frac{d+1}{2})}{(2\pi)^{\frac{d+1}{2}}} mH^d \sum_{l =
0}^{\infty} D^{(d)}_l \Biggl[ {\cal P} \int_0^{\infty} \frac{ds}{s}
\frac{e^{- \gamma s}}{\sin(\frac{s}{2})} \Bigl\{\cos (l+
\frac{d-1}{2})s - \cos(\frac{s}{2}) \Bigr\} \nonumber\\&& + \frac{i}{2} \ln (1 +
\overline{\cal N}_{L}) \Biggr], \label{ds eff act2}
\end{eqnarray}
where ${\cal P}$ denotes the principal value and $\overline{\cal
N}_{L}$ is the mean number (\ref{mean num}) of produced particles with quanta $L$.
The imaginary part is the sum of residues from simple poles at $s = -2 \pi ni$ in the contour integral.
Interestingly, the proper integral converges for each fixed $l$. However, the effective action
(\ref{ds eff act2}) is not finite since the summation over angular momenta is infinite.
The divergent structure of the effective action hides in the angular momentum sum,
which requires a regularization scheme, as will be shown in Sec. 4.

The effective action (\ref{ds eff act2})
has the form similar to the QED effective action in the proper-time formalism by Schwinger \cite{schwinger}.
In the $(d+1)$-dimensional Minkowski spacetime,
a massive scalar with charge $q$ and spin multiplicity 2 in a constant $E$-field has the Bogoliubov coefficients \cite{KLY08,KLY10}
\begin{eqnarray}
\mu_{{\bf k}_{\perp}} = \frac{\sqrt{2 \pi} e^{- i \frac{\pi}{4}} e^{- \pi \frac{m^2+ {\bf k}_{\perp}^2}{4 qE}}}{\Gamma \Bigl( \frac{1}{2} + i \frac{m^2+ {\bf k}_{\perp}^2}{2qE} \Bigr)},
\end{eqnarray}
and thereby the exact one-loop effective action per unit volume and per unit time
\begin{eqnarray}
{\cal L}^{(d+1)}_{\rm eff} (E) = - \frac{qE}{(2 \pi)} \int \frac{d^{d-1}
{\bf k}_{\perp}}{(2 \pi)^{d-1}} \Biggl[ \int_0^{\infty} \frac{ds}{s}
\frac{e^{- \frac{m^2+ {\bf k}_{\perp}^2}{2qE} s}}{\sin(\frac{s}{2})}
 -i \ln \Bigl(1 + \overline{\cal N}_{{\bf k}_{\perp}} \Bigr) \Biggr]. \label{E-eff act}
\end{eqnarray}
Here, ${\bf k}_{\perp}$ is the momentum component transverse to the electric field and
the mean number of produced pairs via Schwinger mechanism is
\begin{eqnarray}
\overline{\cal N}_{{\bf k}_{\perp}} = e^{- \pi \frac{m^2+ {\bf k}_{\perp}^2}{qE}}.
\end{eqnarray}
It is worthy to note the similarity and difference between the dS effective action (\ref{ds eff act2}) and the QED one (\ref{E-eff act}). Quantum fluctuations are angular excitations in the former case whereas they are the transverse motions in the latter case, and $\gamma$ corresponds to $\frac{m^2+ {\bf k}_{\perp}^2}{2qE}$. The function $1/s \sin(\frac{s}{2})$ is characteristic to bosons. However, the divergent structure comes from the infinite sum over angular momenta in the dS effective action while it is the singularity of proper-time integral in the QED effective action, which should be regularized through the vacuum (mass) energy and the charge renormalization, etc. After the momentum integral, the exact one-loop effective action is given by
\begin{eqnarray}
{\cal L}^{(d+1)}_{\rm eff} (E) = - \Bigl(\frac{qE}{2\pi} \Bigr)^{(d+1)/2} {\cal P}
\int_0^{\infty} \frac{ds}{s^{(d+1)/2}}  \frac{e^{ - \frac{m^2}{2qE} s}
}{\sin(\frac{s}{2})} -
\frac{i}{(2\pi)^{d}} \sum_{n=1}^{\infty} \Bigl( \frac{qE}{n} \Bigr)^{(d+1)/2}
(-e^{-\frac{\pi m^2}{qE}})^{n}.
\label{E-eff2}
\end{eqnarray}

A few comments are in order. In all odd dimensions
the effective action (\ref{ds eff act2}) does not have an imaginary part, implying no
particle production, whereas in even dimensions the existence of the imaginary part has been a controversial issue \cite{polyakov,akhmedov}. The effective action (\ref{ds eff act2}), modulo angular momentum summation, has  the imaginary part in any even dimension and thereby the mean number of produced particles, thus resolving the controversial issue. Further, the general relation holds in any dimension, even or odd, between the imaginary part and the mean number
\begin{eqnarray}
2 ({\rm Im} {\cal L}_{\rm eff}) = \sum_{L} \ln (1+ \overline{\cal N}_{L}),
\end{eqnarray}
which is a consequence of the Bogoliubov transformation (\ref{bog tran}) and the definition of the effective action (\ref{eff act}). The dS radiation can be interpreted as Schwinger mechanism with the Unruh temperature $H/(2 \pi)$ in analogy to QED, whose Unruh temperature is the inverse of the period of the Euclidean motion \cite{kim07,hwang-kim}.

\section{Zeta-Function Regularization of Weak-Curvature Expansion}

The effective action (\ref{ds eff act2}) is equivalent to the sum of all Feynman diagrams with one internal loop of a massive scalar interacting with arbitrary number of dS gravitons just as the QED effective action (\ref{E-eff act}) or (\ref{E-eff2}) is the sum of all one-loop diagrams with arbitrary even number of photons. Though at higher energy or large action the quantized scalar field may break down or be one sector of unified theory, in this paper we adopt the exact one-loop effective action somewhat in a literal sense. Then the real part of the effective action (\ref{ds eff act2}) is not renormalized since
the infinite sum over the angular momenta $l$ makes it diverge. So any divergent term from this summation should be properly regularized to yield the renormalized effective action.
In this section we propose the zeta-function regularization since the summation over the angular momenta can be expressed in terms of the Riemann zeta functions or the Hurwitz zeta functions.

Let us first investigate the weak-field expansion of the real part (vacuum polarization) of QED effective action (\ref{E-eff2}), which follows by expanding $1/\sin (\frac{s}{2})$,
\begin{eqnarray}
{\rm Re} ({\cal L}^{(d+1)}_{\rm eff} (E)) &=& - \Bigl(\frac{qE}{4\pi} \Bigr)^{(d+1)/2} \Biggl[2 \Gamma \Bigl(- \frac{d+1}{2} \Bigr) \Bigl(\frac{qE}{m^2} \Bigr)^{-(d+1)/2}  +  \sum_{k = 1}^{\infty} \frac{2^2 (2^{2k-1} -1)}{(2k)!}  \nonumber\\&& \times \vert B_{2k} \vert \Gamma \Bigl(2k - \frac{d+1}{2} \Bigr)  \Bigl(\frac{qE}{m^2} \Bigr)^{2k - (d+1)/2} \Biggr], \label{E-exp}
\end{eqnarray}
where $B_{2k}$ are Bernoulli numbers. Those terms with $2k \leq \frac{d+1}{2}$ are apparently singular in Eq. (\ref{E-eff2}), so these terms should be regularized, for instance, $k = 0$ corresponding to the energy (mass) renormalization and $k =1$ to the charge renormalization, etc.
Hence, the renormalized QED effective action is a series of the terms $(qE)^{(d+1)/2} (\frac{qE}{m^2})^{2 k - (d+1)/2}$ for $2k > \frac{d+1}{2}$ and in $d+1 = 4$, the leading term is $(qE)^{2} (\frac{qE}{m^2})^{2}$, as expected. The analytical continuation of the gamma functions distinguishes odd dimensions $(d~{\rm even~integer})$ from even dimensions $(d~{\rm even~integer})$. The weak-field expansion does not converge due to factorially growing coefficients from symmetric factors for Feynman diagrams. Further, the Borel summation of the non-alternating series (\ref{E-exp}) leads to an imaginary part, explaining Schwinger mechanism in QED regardless of dimensions \cite{das-dunne}.

We turn to the effective action in dS spaces. Similarly, expanding $1/\sin (\frac{s}{2})$, $\cos (l+ \frac{d-1}{2})s$, and $\cos(\frac{s}{2})$, respectively,
we obtain the weak-curvature expansion of the real part of the effective action
\begin{eqnarray}
{\rm Re} ({\cal L}^{(d+1)}_{\rm eff} (H)) &=&  \frac{\Gamma (\frac{d+1}{2})}{(2\pi)^{\frac{d+1}{2}}} mH^d
 \sum_{l
= 0}^{\infty} D^{(d)}_l {\cal P} \int_0^{\infty} ds e^{- \gamma s}
\Biggl[ \frac{2}{s^2} + \sum_{k = 1}^{\infty} \frac{2^{2k-1}
-1}{(2k)!} \vert B_{2k} \vert \Bigl(\frac{s}{2} \Bigr)^{2k-2}
\Biggr] \nonumber\\ && \times \sum_{n =1}^{\infty} \frac{(-1)^n}{(2n)!} \Biggl[ \Bigl(l+
\frac{d-1}{2} \Bigr)^{2n} - \frac{1}{2^{2n}} \Biggr] s^{2n}. \label{vac
pol}
\end{eqnarray}
The proper integral yields a series of $\frac{1}{\gamma^{2n-1}}$. In the large-mass limit
of $m \gg \frac{dH}{2} = \frac{1}{2} \sqrt{\frac{dR}{d+1}}$, the real part of the effective action is a series of $R^{(d+1)/2} (\frac{R}{m^2})^{n-1}$ for $n \geq 1$, and the leading term is $R^{(d+1)/2}$, being $R^2$ in $d+1 = 4$. A closer inspection shows that the scalar curvature $R$ in dS space corresponds to $(qE)^2$ in QED, which suggests the gauge-gravity relation for AdS and $B$-field may possibly hold between a 2n-dimensional dS space and a 4n-dimensional electric field. Now, the infinite sum over $l$ can be regularized through the zeta functions, as will be shown below.
As the degeneracies (\ref{sph har}) of spherical harmonics discriminate the dimensionality
of spacetimes, we separately treat the even and odd dimensions.

\subsection{Even Dimensional dS Space}

In even dimensions, $d = 2p -1,~(p \geq 1)$, the degeneracies take the form
\begin{eqnarray}
D_l^{(2p-1)} = \frac{2}{(2p-2)!} \prod_{q = 0}^{p-2} (x^2 - q^2),
\quad (x = l+ p-1), \label{deg ev}
\end{eqnarray}
which can be written as a polynomial
\begin{eqnarray}
D_l^{(2p-1)} = \frac{2}{(2p-2)!} \sum_{j = 1}^{p-1} C^{(2p-1)}_{2j} x^{2j}.
\end{eqnarray}
Then, the summation over $l$ in Eq. (\ref{vac pol}) becomes Riemann zeta functions \cite{zeta function}, modulo a finite sum,
\begin{eqnarray}
\sum_{j = 1}^{p-1} C^{(2p-1)}_{2j} \sum_{l = 0}^{\infty} (l+ p-1)^{2j+2n} =
\sum_{j = 1}^{p-1} C^{(2p-1)}_{2j}  \zeta (-2j-2n) - \sum_{j = 1}^{p-1} C^{(2p-1)}_{2j} \sum_{l = 0}^{p-2}
l^{2j+2n}. \label{sum even}
\end{eqnarray}
The relations from the polynomial (\ref{deg ev})
\begin{eqnarray}
\sum_{j = 1}^{p-1} C^{(2p-1)}_{2j} l^{2j} = 0, \quad (l = 0, \cdots, p-2), \label{iden1}
\end{eqnarray}
make the second summation vanish, so we obtain
\begin{eqnarray}
\sum_{j = 1}^{p-1} C^{(2p-1)}_{2j} \sum_{l = 0}^{\infty} (l+ p-1)^{2j+2n} =
\sum_{j = 1}^{p-1} C^{(2p-1)}_{2j} \zeta (-2j-2n). \label{sum even2}
\end{eqnarray}

Finally, the weak-curvature expansion of the effective action is expressed in terms of zeta functions as
\begin{eqnarray}
{\rm Re} ({\cal L}^{(2p)}_{\rm eff} (H)) &=&  \frac{2 \Gamma(p)}{2 \pi^p (2p-2)!}
mH^{2p-1} \sum_{j = 1}^{p-1} C^{(2p-1)}_{2j} \sum_{n =1}^{\infty} \frac{(-1)^n}{(2n)!}
\Biggl[\zeta (-2j-2n) - \frac{1}{2^{2n}} \zeta (-2j) \Biggr] \nonumber
\\ && \times   \Biggl[
\frac{2 \Gamma (2n-1)}{\gamma^{2n-1}} + \sum_{k = 1}^{\infty}
\frac{(2^{2k-1} -1)\vert B_{2k} \vert \Gamma (2k+ 2n-1)}{(2k)!
2^{2k-2} \gamma^{2k+ 2n-1}} \Biggr], \label{eff ds even}
\end{eqnarray}
We now employ the zeta-function regularization \cite{hawking} (for review and references, see \cite{EORBZ})
\begin{eqnarray}
\zeta (-2n) = 0, \quad (n = 1, 2, \cdots).
\end{eqnarray}
The zeta-function regularization makes the real part of the effective action zero. Similarly, as the mean number is independent of $l$, the imaginary part of the effective action (\ref{ds eff act2}) becomes
\begin{eqnarray}
{\rm Im} ({\cal L}^{(2p)}_{\rm eff} (H)) &=& \frac{\Gamma (\frac{d+1}{2})}{2 (2\pi)^{\frac{d+1}{2}}} mH^d \ln (1 + \overline{\cal N}) \sum_{l = 0}^{\infty} D^{(2p-1)}_l.
\end{eqnarray}
The summation of angular momenta
\begin{eqnarray}
\sum_{l = 0}^{\infty} D^{(2p-1)}_l = \frac{2}{(2p-2)!} \sum_{j= 1}^{p-1} \Biggl[ C^{(2p-1)}_{2j} \zeta (-2j) - \sum_{l = 0}^{p-2} C^{(2p-1)}_{2j} l^{2j} \Biggr],
\end{eqnarray}
also vanishes due to the zeta-function regularization and the relations (\ref{iden1}). The zeta-function regularization for the angular momenta in Eq. (\ref{ds eff act2}), after expanding only $\cos (l+ \frac{d-1}{2})s$ and $\cos(\frac{s}{2})$ but keeping $1/\sin (\frac{s}{2})$, leads the zero effective action. We may thus conclude that the renormalized effective action in the weak-curvature expansion, both the real and imaginary parts, vanishes for all even dimensions.

\subsection{Odd Dimensional dS Space}

In odd dimensions, $d = 2p,~(p \geq 1)$, we write the degeneracies as
\begin{eqnarray}
D_l^{(2p)} = \frac{2x}{(2p-1)!} \prod_{q = 1}^{p-1} \Biggl[ x^2 -
\Bigl(\frac{2p - (2q+1)}{2} \Bigr)^2 \Biggr], \quad (x = l+ p-1 + \frac{1}{2}),
\label{deg od}
\end{eqnarray}
and expand it as a polynomial
\begin{eqnarray}
D_l^{(2p)} = \frac{2}{(2p-1)!} \sum_{j = 1}^{p} C^{(2p)}_{2j-1} x^{2j-1}.
\end{eqnarray}
Since the following relations hold from the polynomial (\ref{deg od}),
\begin{eqnarray}
\sum_{j = 1}^{p} C^{(2p)}_{2j-1} \Bigl(l + \frac{1}{2} \Bigr)^{2j-1} = 0, \quad (l = 0, \cdots, p-2).
\end{eqnarray}
the sum over $l$ is simplified as Hurwitz zeta functions \cite{zeta function}
\begin{eqnarray}
\sum_{j = 1}^{p} C^{(2p)}_{2j-1} \sum_{l = 0}^{\infty} \Bigl(l+ p-1 +
\frac{1}{2} \Bigr)^{2j+2n-1} = \sum_{j = 1}^{p} C^{(2p)}_{2j-1} \zeta(-2j-2n+1, \frac{1}{2}). \label{sum odd2}
\end{eqnarray}
Therefore, the weak-curvature expansion of the effective action in odd dimensions takes the form
\begin{eqnarray}
{\rm Re} ({\cal L}^{(2p+1)}_{\rm eff} (H)) &=&  \frac{2 \Gamma (\frac{2p+1}{2})}{2 \pi^{\frac{2p+1}{2}} (2p-1)!}
mH^{2p} \sum_{j = 1}^{p} C^{(2p)}_{2j-1} \sum_{n =1}^{\infty} \frac{(-1)^n}{(2n)!}
\nonumber\\ && \times \Biggl[\zeta (-2j-2n +1, \frac{1}{2}) - \frac{1}{2^{2n}} \zeta (-2j
+1, \frac{1}{2}) \Biggr]  \nonumber
\\ && \times \Biggl[ \frac{2 \Gamma
(2n-1)}{\gamma^{2n-1}} + \sum_{k = 1}^{\infty} \frac{(2^{2k-1}
-1)\vert B_{2k} \vert \Gamma (2k+ 2n-1)}{(2k)! 2^{2k-2} \gamma^{2k+
2n-1}} \Biggr]. \label{eff act odd}
\end{eqnarray}
Note that $\zeta (-2j-2n +1, \frac{1}{2})= - B_{2j+ 2n +2} (\frac{1}{2})/(2j+2n+2)$. The imaginary part is identically zero since there is no particle production in odd dimensions as explained in Sec. 3.

\section{Conclusion}

In this paper, employing the in-/out-state formalism, we have found the exact one-loop effective action of a massive scalar in dS space in any dimension and compared it with QED effective action in a constant electric field. The effective action is defined as the $S$-matrix between the in-vacuum and the out-vacuum. The effective action (\ref{ds eff act2}) is the gravity analog of the Heisenberg-Euler effective action (\ref{E-eff act}) or (\ref{E-eff2}) in scalar QED, which is equivalent to summing all one-loop diagrams interacting with arbitrary even number of gravitons or photons. It exhibits nonperturbative aspects, such as the imaginary part in even dimensions, and keeps the covariance since it is expressed entirely in terms of the curvature, just as the QED effective action is gauge invariant. One noticeable point is that the proper integral of the effective action is finite for each angular momentum, in contrast with the QED effective action. However, there occur divergent terms from the infinite sum of angular momenta, which are analogs of the ultraviolet divergent terms in Minkowski spacetimes and also of ultraviolet terms in the momentum integral in the QED effective action (\ref{E-eff act}).

The nonperturbative effective actions (\ref{ds eff act2}) and (\ref{E-eff act}), though unrenormalized,  share some points in common. The quantum fluctuations come from angular excitations in dS space while they come from transverse motions in addition to the acceleration of a charged particle in QED. Thus, the exact one-loop effective action is the sum over angular momenta in dS space and the transverse momenta in QED. The sinusoidal function $1/\sin(\frac{s}{2})$ is the property of bosons. From Eqs. (\ref{ds eff act2}) and (\ref{E-eff act}) the scalar curvature $R$ in dS space corresponds to $(qE)^2$ in QED, and the weak-curvature/-field expansion in Sec. 4 suggests a possible gauge-gravity relation between a $2n$-dimensional dS space and a $4n$-dimensional $E$-field, while the gauge-gravity relation is known between a $2n$-dimensional AdS and a $4n$-dimensional $B$-field in Ref. \cite{basar-dunne}.

The regularization and renormalization for dS spaces, however, differs from that for QED due to the nature of divergence structure. The divergences in dS spaces from the infinite angular momenta are countable and can be expressed in terms of Riemann or Hurwitz zeta functions, whereas those in QED come from the integration over the transverse momenta and are regularized through the vacuum energy (mass) and the charge renormalization, etc.
In the weak-curvature expansion of the dS effective action and in the zeta-function regularization, to our surprise, the renormalized effective action vanishes in all even dimensions. The physical implication is that an even dimensional dS space may not have any quantum hair and thus may be stable at one-loop. However, the effective action has only finite real part in any odd dimension. The difference between the even and odd dimensions is the consequence of the degeneracies of the Laplace operator on $S_d$.

In the zeta-function regularization for the weak-curvature expansion of the dS effective action, we assume arbitrarily large angular excitations. At such higher energy or action, the quantum field theory of a massive scalar should include interactions with other fields or quantum gravity may enter here \cite{tsamis-woodard}. If we adopt the exact one-loop effective action in a literal sense, it is free from quantum corrections at one-loop level, that is, there are remarkable cancelations among quantum corrections. In the case of interactions, however, the result of this paper does not rule out the possibility of the decay of dS spaces due to quantum fluctuations and pair production \cite{myhrvold,BVS,BEM,alvarez-vidal,marolf-morrison}. The stability of interacting fields or two-loop effective action in dS spaces will be addressed in the future.

\section*{Acknowledgments}

The author is grateful to Don~N.~Page for valuable comments, is benefitted
from useful discussions with Andrei O.~Barvinsky, and thanks Bum-Hoon~Lee
and Wei-Tou Ni for useful discussions, Akihiro~Ishibashi,
Hideo~Kodama, and Emil~T.~Akhmedov for useful information.
He also thanks W-Y.~ Pauchy Hwang, Misao Sasaki, Remo Ruffini, and Holger Gies for the warm hospitality at Leung Center for Cosmology and Particle Astrophysics of National Taiwan University, Yukawa Institute for Theoretical Physics of Kyoto University, and International Center for Relativistic Astrophysics Network, where parts of this paper were done, and Theoretisch-Physikalisches Institut, Friedrich-Schiller-Universitat Jena, where this paper was finished.
This work was supported by the Korea Research Council of Fundamental Science and Technology (KRCF).

\appendix


\begin{thebibliography}{99}
\bibitem{zeta function} The Riemann zeta function is
$\zeta (z) = \sum_{k = 1}^{\infty} \frac{1}{ k^z}$ and the Hurwitz
zeta function is $\zeta (z, a) = \sum_{k = 0}^{\infty} \frac{1}{(k+a)^z}$.
The complex Riemann zeta function leads to $\zeta (-2n) = 0$ for all
positive $n$ and $\zeta (0) = -\frac{1}{2}$.
\end{thebibliography}
\end{document}